\documentclass[11pt]{article}

\usepackage{geometry}                
\usepackage{graphicx}
\usepackage{amsmath,amssymb,amsfonts,amsthm,bm}
\usepackage{tikz-cd} 
\usepackage{cite}
\usepackage{authblk} 

\newcommand{\vol}{\bm{\mathrm{{vol}}}}

\newtheorem{prop}{Proposition}

\newtheorem{lemma}{Lemma}

\title{A Hamiltonian model for the macroscopic Maxwell equations using exterior calculus}
\author[1]{William Barham}
\author[2]{Philip J. Morrison}
\author[3,4]{Eric Sonnendr{\"u}cker}
\affil[1]{Oden Institute for Computational Engineering and Sciences, The University of Texas at Austin}
\affil[2]{Department of Physics and Institute for Fusion Studies, The University of Texas at Austin}
\affil[3]{Max-Plank-Institut f{\"u}r Plasmaphysik}
\affil[4]{Technische Universit{\"a}t M{\"u}nchen, Zentrum Mathematik}
\date{\today}                     
\begin{document}

\maketitle

\section*{Abstract}

A Hamiltonian field theory for the macroscopic Maxwell equations with fully general polarization and magnetization is stated in the language of differential forms. The precise procedure for translating the vector calculus formulation into differential forms is discussed in detail. We choose to distinguish between straight and twisted differential forms so that all integrals be taken over densities (i.e. twisted top forms). This ensures that the duality pairings, which are stated as integrals over densities, are orientation independent. The relationship between functional differentiation with respect to vector fields and with respect to differential forms is established using the chain rule. The theory is developed such that the Poisson bracket is metric and orientation independent with all metric dependence contained in the Hamiltonian. As is typically seen in the exterior calculus formulation of Maxwell's equations, the Hodge star operator plays a key role in modeling the constitutive relations. As a demonstration of the kind of constitutive models this theory accommodates, the paper concludes with several examples. 

\section{Introduction}
A host of electromagnetic phenomena occur in polarized and magnetized media. As the rational of introducing polarization and magnetization amounts to the hiding complicated microscopic behavior in constitutive laws, the equations describing electromagnetism in a medium are often called the macroscopic Maxwell equations. Typically, an empirical linear model is used for these constitutive models. However, in many plasma models, it is useful to consider a self consistent model that can account for more complex couplings between the material, e.g. a charged particle model, and the fields. A systematic theory for lifting particle models to kinetic models and the Hamiltonian structure of these lifted models was given in \cite{morrison_gauge_free_lifting}. It has been shown that many kinetic models of interest fit into this framework such as guiding center drift kinetics \cite{morrison_gauge_free_lifting} or gyrokinetics \cite{burby_et_al_gyrokinetics}. Frequently, such models include nonlinear field dependent polarizations. 

Recently, structure preserving discretizations which replicate aspects of the continuous Hamiltonian structure in a discrete system have been a popular area of research \cite{morrison_structure_preserving_algorithms}. In particular, such a discretization was found for the Maxwell-Vlasov equations \cite{GEMPIC}. These structure preserving methods generally rely on the characterization of Maxwell's equations using differential forms (e.g. \cite{hiptmair_maxwell_equations}, \cite{arnold_FEEC}, and \cite{kreeft2011mimetic}). The Maxwell component of the model given in \cite{morrison_gauge_free_lifting} involves a somewhat complicated expression for the relationship between the polarization and magnetization with the electric and magnetic fields through functional derivatives of an energy functional. This adds difficulty when incorporating this model into the geometric language of differential forms necessary for structure preserving discretization. In particular, some care is needed when interpreting functional derivatives with respect to differential forms. This work seeks to provide a sound geometric interpretation of the macroscopic Maxwell equations and their Hamiltonian structure. A structure preserving discretization of the macroscopic Maxwell equations in Hamiltonian form is given in the accompanying paper \cite{dualMimeticMaxwell}.

The macroscopic Maxwell equations obtained from \cite{morrison_gauge_free_lifting} are a Hamiltonian field theory. This paper seeks to express this Hamiltonian field theory in terms of differential forms. Hence, it follows that one must investigate the meaning of variational derivatives with respect to differential forms. In order to properly understand derivatives with respect to differential forms, it is necessary to give some attention to notions of duality in the double de Rham complex. The most obvious notion of duality is the $L^2$ inner product which pairs $k$-forms with themselves. Alternatively, one may use Poincar{\'e} duality to pair differential $k$-forms with orientation dependent twisted $(n-k)$-forms (a pairing which does not utilize a metric). This leads to two alternative identifications of the functional derivative which are Poincar{\'e} duals of each other through the Hodge star operator.

Using the split exterior calculus formalism presented in \cite{eldred_and_bauer}, we obtain a Hamiltonian model for the macroscopic Maxwell equations which is orientation independent and whose Poisson bracket is explicitly metric independent. This manner of expressing a Hamiltonian theory is believed to be advantageous for discretization using the novel split Hamiltonian finite element framework \cite{bauer:shallow_water_split_FEM} or the split mimetic spectral element framework developed in \cite{dualMimeticMaxwell}. The resulting Hamiltonian theory is easily converted into a format based only on the primal complex, but certain features (e.g. explicit metric independence of the Poisson bracket) are lost in this conversion. Treatment of linear constitutive relations in electromagnetism through Poincar{\'e} duality is also found in \cite{hiptmair_maxwell_equations}. Finally, we conclude by considering several motivating examples of constitutive models that that this framework accommodates.

\section{The double de Rham complex and Maxwell's equations}
In this section, we discuss the basic objects of exterior calculus and use them to translate Maxwell's equations from the language of vector calculus to exterior calculus.

\subsection{The macroscopic Maxwell equations}
Maxwell's equations may be written in two equivalent formats:
\begin{equation}
	\begin{aligned}[c]
		\nabla \cdot \bm{D} &= 4 \pi \rho_f, \\
		\nabla \cdot \bm{B} &= 0, \\
		- c \nabla \times \bm{E} &= \frac{\partial \bm{B}}{\partial t}, \\
		c \nabla \times \bm{H} &= \frac{\partial \bm{D}}{\partial t} + 4 \pi \bm{J}_f,
	\end{aligned}
	\qquad \text{or} \qquad
	\begin{aligned}[c]
		\nabla \cdot \bm{E} &= 4 \pi \rho \\
		\nabla \cdot \bm{B} &= 0 \\
		- c \nabla \times \bm{E} &= \frac{\partial \bm{B}}{\partial t} \\
		c \nabla \times \bm{B} &= \frac{\partial \bm{E}}{\partial t} + 4 \pi \bm{J}.
	\end{aligned}
\end{equation}
The left column is frequently called the macroscopic Maxwell equations, and the right column are just the standard Maxwell equations. One relates the displacement field, $\bm{D}$, and the magnetic field intensity, $\bm{H}$, with the electric field, $\bm{E}$, and magnetic field, $\bm{B}$, by
\begin{equation}
	\bm{D} = \bm{E} + 4 \pi \bm{P} \quad \text{and} \quad \bm{H} = \bm{B} - 4 \pi \bm{M}
\end{equation}
where $\bm{P}$ is the polarization and $\bm{M}$ is the magnetization. Moreover, $\rho_f$ and $\bm{J}_f$, the free charge and free current densities respectively, are related to $\rho$ and $\bm{J}$ by
\begin{equation}
	\rho = \rho_b + \rho_f \quad \text{and} \quad \bm{J} = \bm{J}_b + \bm{J}_f
\end{equation}
where
\begin{equation}
	\rho_b = - \nabla \cdot \bm{P} \quad \text{and} \quad
	\bm{J}_b = \bm{J}_p + \bm{J}_m = \frac{\partial \bm{P}}{\partial t} + c \nabla \times \bm{M}.
\end{equation}
It is straightforward to show that the two forms of Maxwell's equations are equivalent using these definitions. The utility of the macroscopic Maxwell equations is that the terms which couple Maxwell's equations to a matter model, $\rho_f$ and $\bm{J}_f$, are prescribed by freely propagating charged particles rather than the, possibly complicated, bound charges and currents in the medium. These microscopic features bound to the medium are instead encoded in constitutive relations for $\bm{P}$ and $\bm{M}$. 

This paper is concerned with combining two distinct modeling frameworks to study this system. First, we translate this system into the language of split exterior calculus \cite{eldred_and_bauer}. Second, we translate the Hamiltonian structure of this model, discovered in \cite{morrison_gauge_free_lifting}, into the language of split exterior calculus. We shall see that this approach helps to elucidate the rich structure of these equations. 

\subsection{The double de Rham complex}
The modeling framework employed in this paper utilizing the double de Rham complex is inspired by the split exterior calculus framework of \cite{eldred_and_bauer}. Discussion of the differential geometric framework is kept to a minimum here, and the reader is referred to the aforementioned split exterior calculus paper and to refereces such as \cite{frankel_2011} and \cite{marsden_and_ratiu}. All notation used herein will either be standard as found in the previous references, or shall use notation defined below. 

Let $(\Omega, g)$ be a Riemannian manifold of dimension $n$. Throughout this paper, we assume either periodic, or homogeneous boundary conditions. Let $\{ (\Lambda^k, \mathsf{d}_k) \}_{k=0}^n$ be the vector spaces of differential forms on $\Omega$. We may define a second complex, $\{ (\tilde{\Lambda}^k, \tilde{\mathsf{d}}_k) \}_{k=0}^n$, called the complex of twisted differential forms. This dual complex differs from the first in that twisted forms change sign under orientation changing transformations. The two complexes are related to each other through the Hodge star operator. Diagrammatically, this is given by
\begin{equation}
	\begin{tikzcd}
		\cdots \arrow{r} & \Lambda^k \arrow{r}{ \mathsf{d}_k } \arrow{d}{\star} & \Lambda^{k+1} \arrow{r} \arrow{d}{\star} & \cdots \\
		\cdots & \tilde{\Lambda}^{n-k} \arrow{u}  \arrow{l} & \tilde{\Lambda}^{n-(k+1)} \arrow{l}{ \tilde{\mathsf{d}}_{n-(k+1)}} \arrow{u} & \cdots \arrow{l} 
	\end{tikzcd}
\end{equation}

While the primal de Rham complex of straight differential forms, $\{\Lambda^k\}_{k=0}^n$, is a familiar object in differential geometry, the dual complex of twisted differential forms, $\{ \tilde{\Lambda}^k \}_{k=0}^n$ would benefit from further discussion. Let $\{ \partial_i \}_{i=1}^n$ be a coordinate system on the $T\Omega$, the tangent space of $\Omega$. We define an operator $o(\bm{\partial})$, which encodes the orientation of the manifold, such that it transforms as
\begin{equation}
	o(T\bm{\partial}) = o(T\partial_1, \hdots, T\partial_n) = \text{sign}(\det(T)) o(\partial_1, \hdots, \partial_n) = \text{sign}(\det(T)) o(\bm{\partial}).
\end{equation}
We may consider twisted differential forms to be straight forms which have been multiplied by this operator. Because the orientation rarely plays a role in the algebraic manipulation of differential forms, for conciseness, we suppress any further use of this operator to the background. 

We can see that the sign of a twisted form depends on the parity of the coordinate system of the ambient space. Therefore, integration over a set $U \subset \Omega$ of a twisted $n$-form will return a value which is independent of the orientation of space. Hence, densities are modeled as twisted forms. As an example, the volume form, that $n$-form such that $\vol(\partial_1, \hdots, \partial_n) = 1$ for any orthonormal frame $\{ \partial_i \}_{i=1}^n$, is a twisted form. We shall typically denote a straight $k$-form as $\omega^k$ and a twisted $k$-form as $\tilde{\omega}^k$. 

An important feature of the split exterior calculus is two distinct notions of duality on the double de Rham complex. First, we have the standard $L^2$ inner product, $( \cdot, \cdot ): V^k \times V^k \to \mathbb{R}$, which is defined
\begin{equation}
	( \omega^k, \eta^k) = \int_\Omega g_x( \omega^k, \eta^k) \vol^n
\end{equation}
where $g_x$ is the pointwise inner product on $k$-forms induced by the Riemannian metric. The second notion of duality is Poincar{\'e} duality, $\langle \cdot, \cdot \rangle: V^k \times \tilde{V}^{n-k} \to \mathbb{R}$, which is defined
\begin{equation}
	\left\langle \omega^k, \tilde{\eta}^{n-k} \right\rangle = \int_\Omega \omega^k \wedge \tilde{\eta}^{n-k}.
\end{equation}
The $L^2$ inner product, because of its dependence on the Riemannian metric and volume form, is a metric dependent quantity. On the other hand, the Poincar{\'e} duality pairing, built from the wedge product structure alone, is purely topological. Moreover, as both duality pairings are expressed as an integral of a twisted $n$-form, they are independent of the orientation of the coordinate system. 

Finally, the Hodge star operator $\star: V^k \to \tilde{V}^{n-k}$ is defined such that
\begin{equation}
	\left( \omega^k, \eta^k \right) = \left\langle \omega^k, \star \eta^k \right\rangle.
\end{equation}
Note that the Hodge star is not a single operator, but rather a family of operators. A more precise notation might be $\star_{n-k,k}: V^k \to \tilde{V}^{n-k}$, however we opt for the more concise notation where confusion is unlikely. These notions of duality are essential for discussing the Hamiltonian structure of the macroscopic Maxwell equations. 

\subsection{The correspondence between vector calculus and exterior calculus}
We denote the space of vector fields on $\Omega$ by $\mathfrak{X}$. Vector fields and $1$-forms are isomorphic to each other. The index lowering operator or flat operator $( \cdot )^\flat : \mathfrak{X} \to \Lambda^1$ is defined by
\begin{equation}
	v^1 = g( \cdot, V) = g_{ij} V^j \mathsf{d} x^i := V^\flat.
\end{equation}
The inverse operation is called the index raising operator or sharp operator $(\cdot)^\sharp: \Lambda^1 \to \mathfrak{X}$:
\begin{equation}
	V = g^*(\cdot, v^1) = g^{ij} v^1_j \frac{\partial}{\partial x^i} := (v^1)^\sharp
\end{equation}
where $g^*$ is the dual metric on the covectors. We may likewise define an isomorphism between vector fields and twisted $(n-1)$-forms. We define $\textbf{i}_{(\cdot)} \vol^n : \mathfrak{X} \to \tilde{\Lambda}^{n-1}$ by
\begin{equation}
	\tilde{v}^{n-1} = \textbf{i}_V \vol^n = \sum_{i} V^i \sqrt{\det(g)} \mathsf{d} x^1 \wedge \hdots \wedge \widehat{\mathsf{d} x^i} \wedge \hdots \wedge \mathsf{d} x^n
\end{equation}
where the hat symbol means omission of ``$\mathsf{d} x^i$" from the wedge product and $\textbf{i}_{V} \alpha$ is the interior product. It is possible to show that $\textbf{i}_V \vol^n = \star V^\flat$. Hence, the inverse operation is given by
\begin{equation}
	V = \left( \star \tilde{v}^{n-1} \right)^\sharp.
\end{equation}
Many physically significant quantities are orientation dependent. The magnetic field and vorticity of a fluid are notable examples of so called pseudovectors. If $V$ is a pseudovector, then one can see that $\tilde{v}^1 = V^\flat$ is twisted while $v^{n-1} = \textbf{i}_V \vol^n$ is straight. Hence, pseudovectors are naturally identified with straight $(n-1)$-forms. 

It is helpful to establish the following adjoint relationships between the isomorphisms we just defined. We define the $L^2$ inner product of vector fields to be
\begin{equation*}
	(W,V) = \int_M W \cdot V \vol^n
\end{equation*}
where $W \cdot V = W^i g_{ij} V^j$.

\begin{prop}
With respect to the duality pairing between $\Lambda^1$ and $\tilde{\Lambda}^{n-1}$ and the $L^2$-inner product on $\mathfrak{X}$,
\begin{equation} \label{adjoint_of_flat}
	 \left( (\cdot)^\flat \right)^* 
	 = \left( \normalfont\textbf{i}_{(\cdot)} \vol^n \right)^{-1} \quad \text{and} \quad \left( ( \cdot )^\sharp \right)^* 
	 = \normalfont\textbf{i}_{(\cdot)} \vol^n.
\end{equation}
\end{prop}

\noindent \textit{Proof:} Consider $v^1 = V^\flat$ and $\tilde{w}^{n-1} = \textbf{i}_W \vol^n$. Then
\begin{align*}
	\left\langle \tilde{w}^{n-1}, v^1 \right\rangle 
	= \int_M \tilde{w}^{n-1} \wedge v^1 
	= \int_M \textbf{i}_W \vol^n \wedge V^\flat 
	= \int_M W \cdot V \ \vol^n = (W, V).
\end{align*}
Therefore, since $\tilde{w}^{n-1} = \star W^\flat = \textbf{i}_W \vol^n \iff W = ( \star \tilde{w}^{n-1})^\sharp$, 
\begin{equation*}
	\left\langle \tilde{w}^{n-1}, V^\flat \right\rangle
	= \left( ( \star \tilde{w}^{n-1} )^\sharp, V \right) 
	= \left( \left( \textbf{i}_{(\cdot)} \vol^n \right)^{-1} \tilde{w}^{n-1}, V \right)
\end{equation*}
and similarly
\begin{equation*}
	\left\langle v^1, \textbf{i}_W \vol^n \right\rangle_{\Lambda^1, \tilde{\Lambda}^{n-1}} = \left( (v^1)^\sharp, W \right)_{L^2(M)}.
\end{equation*}
\qed

\subsection{Vector calculus in $\mathbb{R}^3$ and differential forms} \label{section:vector_identities}
We now turn our attention to vector calculus in $\mathbb{R}^3$. Exterior differentiation is inherently metric independent whereas it is well known that vector calculus operators such as the gradient, divergence, and curl depend on one's coordinate system -- often in a very complicated manner. By defining a differential form related to a vectorial quantity, one may obtain simpler expressions for their derivatives which disentangle the metric dependent and independent portions of the operation. The following definitions allow us to reconstruct the standard vector calculus operations from the exterior derivative. 

If $f: \Omega \to \mathbb{R}$ be a scalar field on a Riemannian manifold, its gradient and exterior derivative are related to each other via
\begin{equation}
	\mathsf{d} f = (\nabla f)^\flat \iff \nabla f = ( \mathsf{d} f )^\sharp.
\end{equation}
Let $V$ be a vector field and $v^1 = V^\flat$. Then the curl of $V$ is defined by
\begin{equation} \label{curl_differential_forms}
	\mathsf{d} v^1 = \normalfont\textbf{i}_{\nabla \times V} \vol^3 = \star ( \nabla \times V)^\flat
	\iff
	\nabla \times V = (\star ( \mathsf{d} v^1 ))^\sharp.
\end{equation}
Letting $\tilde{v}^{2} = \textbf{i}_V \vol^3$, then the divergence is defined to be
\begin{equation}
	\mathsf{d} \tilde{v}^2 
	= \mathsf{d} \textbf{i}_V \vol^3
	= \pounds_V \vol^3
	= (\nabla \cdot V) \vol^3
	= \star (\nabla \cdot V)
	\iff
	\nabla \cdot V = \star \mathsf{d} \tilde{v}^2
\end{equation}
where $\pounds_V$ is the Lie derivative. The divergence of a vector field physically corresponds with the the change in the volume element as it is dragged by that vector field. It is relatively straightforward to show that these formulas obtain the correct expressions for the gradient, divergence, and curl in Euclidean space and standard curvilinear coordinate systems (e.g. cylindrical and spherical coordinates). 

We next establish a relationship between the wedge product of differential forms and the typical vector and scalar products in $\mathbb{R}^3$.
\begin{prop}
Let $U,V,W \in \mathfrak{X}$ and let $u^1, v^1, w^1$ be their corresponding $1$-forms. Then
\begin{equation}
	\star (u^1 \wedge v^1 \wedge w^1) = \vol^3(U,V,W).
\end{equation}
\end{prop}
\noindent \textit{Proof:} 
\begin{align*}
	u^1 \wedge v^1 \wedge w^1 &= \sum_{ijk} (u_i \mathsf{d} x^i) \wedge (v_j \mathsf{d} x^j) \wedge  (w_k \mathsf{d} x^k) 
						      = \sum_{ijk} u_i v_j w_k \mathsf{d} x^i \wedge \mathsf{d} x^j \wedge \mathsf{d} x^k \\
						    \implies \star (u^1 \wedge v^1 \wedge w^1) 
						    &= \sqrt{\det(g)} \sum_{i < j < k} U^i V^j W^k \epsilon_{ijk}
						    = \vol^3(U,V,W).
\end{align*}
\qed
\\
\noindent Hence, the triple wedge product of three $1$-forms measures the volume of the parallelepiped spanned by their vector proxies. 

\begin{prop} \label{dot_prod_def}
$u^1 \wedge \tilde{v}^2 = U \cdot V \vol^3$.
\end{prop}
\noindent \textit{Proof:} $u^1 \wedge \tilde{v}^2 = u^1 \wedge \textbf{i}_V \vol^3 = \textbf{i}_V \vol^3 \wedge u^1 = \textbf{i}_V ( \vol^3 \wedge u^1 ) + \vol^3 \wedge \textbf{i}_V u^1 = U \cdot V \vol^3$. \qed

\begin{prop}
$u^1 \wedge v^1 = \normalfont\textbf{i}_{U \times V} \vol^3$.
\end{prop}
\noindent \textit{Proof:} A geometrically intuitive definition of the cross product is that $U \times V$ is the unique vector such that $\forall W \in \mathfrak{X}$, 
\begin{equation*}
	U \times V \cdot W = \vol^3(U, V, W).
\end{equation*}
But we already know that
\begin{equation*}
	\vol^3(U, V, W) =  \star u^1 \wedge v^1 \wedge w^1
	\iff
	u^1 \wedge v^1 \wedge w^1 = U \times V \cdot W \vol^3= \textbf{i}_{U \times V} \vol^3 \wedge w^1.
\end{equation*}
according to proposition \ref{dot_prod_def}. Hence,
\begin{equation*}
	\left( u^1 \wedge v^1 - \textbf{i}_{U \times V} \vol^3 \right) \wedge w^1 = 0 \quad \forall w^1 \in \Lambda^1 \iff u^1 \wedge v^1 = \textbf{i}_{U \times V} \vol^3.
\end{equation*}
\qed \\
\noindent This implies that if $U$ and $V$ are standard vectors, then $U \times V$ is a pseudovector. 

\subsection{The macroscopic Maxwell equations in the language of split exterior calculus} \label{sec:macro_maxwell_1}
We now have the machinery to directly translate the standard vector calculus formulation of Maxwell's equations into the language of split exterior calculus. Let
\begin{equation}
	\bm{e}^1 = \bm{E}^\flat, 
	\quad \bm{b}^2 = \textbf{i}_{\bm{B}} \vol^3, 
	\quad \tilde{\bm{d}}^2 =  \textbf{i}_{\bm{D}} \vol^3,
	 \quad \tilde{\bm{h}}^1 = \bm{H}^\flat, 
	 \quad \tilde{\rho} = \star \rho, 
	 \quad \text{and} \quad 
	 \tilde{\bm{j}} = \textbf{i}_{\bm{J}} \vol^3.
\end{equation}
If we transform the equations by applying the appropriate isomorphisms as follows:
\begin{equation*}
	\begin{split}
		\star \left( c \nabla \times \bm{E} - \frac{\partial \bm{B}}{\partial t} \right)^\flat &= 0 \\
		\star (\nabla \cdot \bm{B} ) &= 0
	\end{split}
	\hspace{4em}
	\begin{split}
		\star \left( c \nabla \times \bm{H} - \frac{\partial \bm{D}}{\partial t} - 4 \pi \bm{J}_f \right)^\flat &= 0 \\
		\star ( \nabla \cdot \bm{D} - 4 \pi \rho_f ) &= 0.
	\end{split}
\end{equation*}
Then, we may write the equations in either of two different formats:
\begin{equation} \label{eqn:split_maxwell}
	\begin{aligned}[c]
		\mathsf{d} \tilde{\bm{d}}^2 &= 4 \pi \tilde{\rho}^3_f, \\
		\mathsf{d} \bm{b}^2 &= 0, \\
		-c \mathsf{d} \bm{e}^1 &= \frac{\partial \bm{b}^2}{\partial t}, \\
		c \mathsf{d} \tilde{\bm{h}}^1 &= \frac{\partial \tilde{\bm{d}}^2}{\partial t} + 4 \pi \tilde{\bm{j}}^2_f,
	\end{aligned}
	\qquad \text{or} \qquad
	\begin{aligned}[c]
		\mathsf{d} \star \bm{e}^1 &= 4 \pi \tilde{\rho}^3 \\
		\mathsf{d} \bm{b}^2 &= 0 \\
		-c \mathsf{d} \bm{e}^1 &= \frac{\partial \bm{b}^2}{\partial t} \\
		c \mathsf{d} \star \bm{b}^2 &= \frac{\partial \star \bm{e}^1}{\partial t} + 4 \pi \tilde{\bm{j}}^2.
	\end{aligned}
\end{equation}
The displacement field twisted $2$-form and magnetic intensity field twisted $1$-form, $\tilde{\bm{d}}^2$ and $\tilde{\bm{h}}^1$ respectively, are related to the electric field $1$-form, $\bm{e}^1$, and magnetic field $2$-form, $\bm{b}^2$, by
\begin{equation}
	\tilde{\bm{d}}^2 =  \tilde{\star}_1 \bm{e}^1 + 4 \pi \tilde{\bm{p}}^2 \quad \text{and} \quad \tilde{\bm{h}}^1 = \tilde{\star}_2 \bm{b}^2 - 4 \pi \tilde{\bm{m}}^1 
\end{equation}
where $\tilde{\bm{p}}^2 = \textbf{i}_{\bm{P}} \vol^3$ is the polarization twisted $2$-form, and $\tilde{\bm{m}}^1 = \bm{M}^\flat$ is the magnetization twisted $1$-form. Moreover, the free charge twisted $3$-form and free current twisted $2$-form, $\tilde{\rho}^3_f$ and $\tilde{\bm{j}}^2_f$ respectively, are related to $\tilde{\rho}^3$ and $\tilde{\bm{j}}^2$ by
\begin{equation*}
	\tilde{\rho}^3 = \tilde{\rho}^3_f + \tilde{\rho}^3_b \quad \text{and} \quad \tilde{\bm{j}}^2 = \tilde{\bm{j}}^2_f + \tilde{\bm{j}}^2_b
\end{equation*}
where
\begin{equation*}
	\tilde{\rho}^3_b = - \mathsf{d} \tilde{\bm{p}}^2 \quad \text{and} \quad \tilde{\bm{j}}^2_b = \tilde{\bm{j}}^2_p + \tilde{\bm{j}}^2_m = \frac{\partial \tilde{\bm{p}}^2}{\partial t} + c \mathsf{d} \tilde{\bm{m}}^1.
\end{equation*}
Straightforward algebraic manipulation shows that the two columns of Maxwell's equations are equivalent. The left column is the split exterior calculus formulation of the macroscopic Maxwell equations. The remainder of this paper is dedicated to translating the Hamiltonian structure of the above system, discovered in \cite{morrison_gauge_free_lifting}, into the language of split exterior calculus. 

\section{Variational derivatives with respect to differential forms}
As Hamiltonian field theories are formulated via the calculus of variations, it is necessary to briefly consider the calculus of variations with respect to differential forms. Moreover, as we shall later see, it is necessary to study not only the first variation, but also more complex variational derivatives in order to properly contextualize our Hamiltonian model of the macroscopic Maxwell equations. 

\subsection{The first variation}

Let $K: \Lambda^k \to \mathbb{R}$. We may define a Fr{\'e}chet derivative of this functional in the usual manner
\begin{equation}
	\left| K [\omega + \eta] - K[\omega] - D K[\omega] \eta \right| = O( \| \eta \| ).
\end{equation}
We shall not worry about functional analytic rigor here, but only the formal correctness of our expressions. Note that $DK[\omega] \in (\Lambda^k)^*$, the dual space to the $k$-forms. In order to identify the functional derivative as an element of one of the vector spaces in the double de Rham complex, we must utilize the previously defined duality pairings. Using the $L^2$ and Poincar{\'e} duality structures on the double de Rham complex, we may define two varieties of functional derivatives:
\begin{equation}
	D K[\omega] \eta 
	= \left( \eta, \frac{\delta K}{\delta \omega} \right)
	= \left\langle \eta, \frac{\tilde{\delta} K}{\delta \omega} \right\rangle.
\end{equation}
One can see that 
\begin{equation*}
	\frac{\delta K}{\delta \omega} \in \Lambda^k, \quad \frac{\tilde{\delta} K}{\delta \omega} \in \tilde{\Lambda}^{n-k}, \quad \text{and} \quad \frac{\tilde{\delta} K}{\delta \omega} = \star \frac{\delta K}{\delta \omega}.
\end{equation*}
Notice, $\tilde{\delta} K/ \delta \omega$ is a twisted form while $\delta K/\delta \omega$ is straight. For this reason, we call $\tilde{\delta} K/ \delta \omega$ a twisted functional derivative \cite{eldred_and_bauer}. 

The same procedure may be done for a functional which takes a twisted form as an argument: $K: \tilde{\Lambda}^k \to \mathbb{R}$. In this case, the twisted functional derivative yields a straight form and the straight functional derivative yields a twisted form:
\begin{equation*}
	\frac{\delta K}{\delta \tilde{\omega}} \in \tilde{\Lambda}^k, \quad \frac{\tilde{\delta} K}{\delta \tilde{\omega}} \in \Lambda^{n-k}, \quad \text{and} \quad \frac{\tilde{\delta} K}{\delta \tilde{\omega} } = \star \frac{\delta K}{\delta \tilde{\omega}}.
\end{equation*}
Finally, if we let $\tilde{\omega} = \tilde{\star} \omega \in \tilde{\Lambda}^{n-k}$, then the chain rule implies that $\tilde{\delta} K/ \delta \tilde{\omega} = \delta K/ \delta \omega$. 

The following proposition will be needed to establish the existence of Casimir invariants of the macroscopic Maxwell equations.
\begin{prop}
Let $k \leq n - 1$ and let $\omega \in H^1 \Lambda^k(\Omega)$. Suppose $K[\omega] = \hat{K}[\mathsf{d} \omega]$. Then
\begin{equation} \label{chain_rule_ext_deriv}
	\frac{\tilde{\delta} K}{\delta \omega} = (-1)^{n-k} \mathsf{d} \frac{\tilde{\delta} \hat{K}}{\delta \mathsf{d} \omega}.
\end{equation}
Therefore, $\tilde{\delta} K/ \delta \omega$ is an exact differential form.
\end{prop}
\noindent \textit{Proof:}
The Leibniz rule and chain rule imply that $\forall \eta \in H^1 \Lambda^k(\Omega)$, 
\begin{align*}
	\left\langle \eta, \frac{\tilde{\delta} K}{\delta \omega} \right\rangle 
	&= \left\langle \mathsf{d} \eta,\frac{\tilde{\delta} \hat{K}}{\delta \mathsf{d} \omega} \right\rangle \\
	&= \left\langle \eta, (-1)^{n-k} \mathsf{d} \frac{\tilde{\delta} \hat{K}}{\delta \mathsf{d} \omega} \right\rangle + \int_{\partial \Omega} \text{tr}_{\partial \Omega} \left( \eta \wedge \frac{\tilde{\delta} \hat{K}}{\delta \mathsf{d} \omega} \right)
\end{align*}
where $\text{tr}_{\partial \Omega}: \Lambda^{n-1}(\Omega) \to \Lambda^{n-1}(\partial \Omega)$ is the trace operator \cite{arnold_FEEC}. With enough regularity, this map is just restriction to the boundary. If we assume homogeneous boundary conditions, we find that
\begin{equation*}
	\frac{\tilde{\delta} K}{\delta \omega} = (-1)^{n-k} \mathsf{d} \frac{\tilde{\delta} \hat{K}}{\delta \mathsf{d} \omega}. 
\end{equation*}
\qed

Because of the isomorphisms relating vectors with $1$-forms and $(n-1)$-forms, we can easily establish a relationship between functional derivatives with respect to vector fields and functional derivatives with respect to differential forms. 

\begin{prop}
Let $\bm{V} \in \mathfrak{X}$ and $K: \mathfrak{X} \to \mathbb{R}$. Define the variational derivative with respect to a vector field:
\begin{equation*}
	DK[\bm{V}] \delta \bm{V} = \int_\Omega \left( \frac{\delta K}{\delta \bm{V}} \cdot \delta \bm{V} \right) \mathsf{d} \bm{x}.
\end{equation*}
Now, let $v^1 = \bm{V}^\flat$ and $\tilde{v}^{n-1} = \textbf{i}_{\bm{V}} \vol^n$. Moreover, let $\hat{K}_1[v^1] = \hat{K}_2[\tilde{v}^{n-1}] = K[ \bm{V}]$. Then 
\begin{equation}
	\frac{\tilde{\delta} \hat{K}_1}{\delta v^1} = \normalfont\textbf{i}_{\delta K/\delta \bm{V}} \vol^n \quad \text{and} \quad \frac{\tilde{\delta} \hat{K}_2}{\delta \tilde{v}^{n-1}} = \left( \frac{\delta K}{\delta \bm{V}} \right)^\flat. 
\end{equation}
\end{prop}
\noindent \textit{Proof:} Using the chain rule and the adjoint identities for the musical isomorphisms, we find
\begin{equation*}
	\frac{\tilde{\delta} \hat{K}_1}{\delta v^1} = \left( \left( (\cdot)^\flat \right)^* \right)^{-1} \frac{\delta K}{\delta \bm{V}} = \textbf{i}_{\delta K/\delta \bm{V}} \vol^n 
	\quad \text{and} \quad 
	\frac{\tilde{\delta} \hat{K}_2}{\delta \tilde{v}^{n-1}} = \left( \left( \textbf{i}_{(\cdot)} \vol^n \right)^* \right)^{-1} \frac{\delta K}{\delta \bm{V}} = \left( \frac{\delta K}{\delta \bm{V}} \right)^\flat
\end{equation*}
\qed

\subsection{Higher order functional derivatives}
Consider a map $\phi: \Lambda^k \to \Lambda^l$. The Fr{\'e}chet derivative of $\phi$ is defined by
\begin{equation}
	\left\| \phi [\omega + \eta] - \phi [\omega] - D \phi [\omega] \eta \right\| = O \left(\| \eta \| \right).
\end{equation}
We identify this via either of the two duality pairings:
\begin{equation*}
	D \phi [\omega] (\eta, \xi) = \left( \eta, \frac{\delta \phi}{\delta \omega} \xi \right) = \left\langle \eta, \frac{\tilde{\delta} \phi}{\delta \omega} \xi \right\rangle.
\end{equation*}
This is the general pattern for the higher order derivatives considered in this section.

By taking the Fr{\'e}chet derivative of $D K [\omega] \in (\Lambda^k)^*$, one obtains the second variation:
\begin{equation*}
	\left| D K [\omega + \xi] \eta - D K [\omega] \eta - D^2 K [\omega] ( \eta, \xi) \right| 
	= O \left( \| \eta \| \| \xi \| \right).
\end{equation*}
where $D^2 K [\omega] (\cdot, \cdot)$ is a symmetric bilinear form. Using our two notions of duality, we may write
\begin{equation}
	D^2 K [\omega] ( \eta, \xi) 
	= \left( \eta, \frac{\delta^2 K}{\delta \omega \delta \omega} \xi \right)
	= \left\langle \eta, \frac{\tilde{\delta}^2 K}{\delta \omega \delta \omega} \xi \right\rangle.
\end{equation}
From this definition, it is clear that
\begin{equation*}
	\frac{\delta^2 K}{\delta \omega \delta \omega} : \Lambda^k \to \Lambda^k \quad \text{and} \quad  \frac{\tilde{\delta}^2 K}{\delta \omega \delta \omega}: \Lambda^k \to \tilde{\Lambda}^{n-k}.
\end{equation*}
We can relate these two versions of the second variation via
\begin{equation*}
	\frac{\tilde{\delta}^2 K}{\delta \omega \delta \omega} = \star \frac{\delta^2 K}{\delta \omega \delta \omega}.
\end{equation*}
As previously noted, the second variation is a symmetric operator: 
\begin{equation*}
	D^2 K [\omega](\eta, \xi) = D^2 K [\omega](\xi, \eta).
\end{equation*}
Hence, it follows that
\begin{equation*}
	\left( \eta, \frac{\delta^2 K}{\delta \omega \delta \omega} \xi \right) = \left( \frac{\delta^2 K}{\delta \omega \delta \omega} \eta, \xi \right) \\
	\quad \text{and} \quad 
	\left\langle \eta, \frac{\tilde{\delta}^2 K}{\delta \omega \delta \omega} \xi \right\rangle = \left\langle \xi, \frac{\tilde{\delta}^2 K}{\delta \omega \delta \omega} \eta \right\rangle.
\end{equation*}

If $K: \Lambda^k \times \Lambda^l \to \mathbb{R}$, then we may introduce the notion of partial functional derivatives:
\begin{equation*}
	\left| K[\omega + \nu, \eta] - K[\omega, \eta] - D_1 K[\omega, \eta] \nu \right| = O \left( \| \nu \| \right)
\end{equation*}
and 
\begin{equation*}
	\left| K[\omega, \eta + \xi] - K[\omega, \eta] - D_2 K[\omega, \eta] \xi \right| = O \left( \| \xi \| \right).
\end{equation*}
Notice, we indicate the partial functional derivatives with subscripts to indicate with respect to which argument we are varying. From here, we may define the total functional derivative:
\begin{equation}
	D K[\omega, \eta](\nu, \xi) = D_1 K[\omega, \eta] \nu + D_2 K[\omega, \eta] \xi,
\end{equation}
and we may identify these using either of the two duality pairings:
\begin{equation*}
	D K[\omega, \eta](\nu, \xi) 
		= \left( \nu, \frac{\delta K}{\delta \omega} \right) + \left( \xi, \frac{\delta K}{\delta \eta} \right)
		= \left\langle \nu, \frac{ \tilde{\delta} K}{\delta \omega} \right\rangle + \left\langle \xi, \frac{ \tilde{\delta} K}{\delta \eta} \right\rangle.
\end{equation*}
Hence, the total functional derivative is just the standard Fr{\'e}chet derivative on a tensor product space. 

Finally, we have mixed second partial derivatives. For example,
\begin{equation*} 
	\left| D_1 K [\omega, \eta + \xi] \nu - D_1 K [\omega, \eta] \nu - D^2_{12} K[\omega, \eta] ( \nu, \xi) \right|
	= O \left( \| \nu \| \| \xi \| \right).
\end{equation*}
As before, there are two natural identifications of the second functional derivative:
\begin{equation*}
	D^2_{12} K(\omega, \eta) ( \nu, \xi) 
	= \left( \nu, \frac{\delta^2 K}{\delta \eta \delta \omega} \xi \right) 
	= \left\langle \nu, \frac{\tilde{\delta}^2 K}{\delta \eta \delta \omega} \xi \right\rangle.
\end{equation*}
The mixed second partial functional derivative is symmetric with respect to interchange of order of differentiation:
\begin{equation*}
	D^2_{12} K = D^2_{21} K.
\end{equation*}

Suppose that $\phi: \mathfrak{X} \to \mathfrak{X}$ and let $\hat{\phi}: \Lambda^1 \to \tilde{\Lambda}^{n-1}$ such that $\phi[\bm{V}] = (\star \hat{\phi}[ \bm{V}^\flat])^\sharp$. Then letting $w^1 = \bm{W}^\flat$, $v^1 = \bm{V}^\flat$, and $\xi^1 = \bm{X}^\flat$, we find
\begin{equation*}
	D \hat{\phi}[v^1] (w^1, \xi^1) 
    	= \left\langle w^1, \frac{\tilde{\delta} \hat{\phi}}{\delta v^1} \xi^1 \right\rangle 
	= \int_\Omega w^1 \wedge \frac{\tilde{\delta} \hat{\phi}}{\delta v^1} \xi^1 
	= \int_\Omega \bm{W} \cdot \left( \star \frac{\tilde{\delta} \hat{\phi}}{\delta v^1} \xi^1 \right)^\sharp \vol^n.
\end{equation*}
Therefore,
\begin{equation} \label{eq:deriv_of_map}
	D \phi[\bm{V}](\bm{W}, \bm{X}) = \left( \bm{W}, \frac{\delta \phi}{\delta \bm{V}} \bm{X} \right) \implies \frac{\delta \phi}{\delta \bm{V}} \bm{X} = \left( \star \frac{\tilde{\delta} \hat{\phi}}{\delta v^1} \xi^1 \right)^\sharp.
\end{equation}
Similar results hold if we identify the domain or codomain of $\hat{\phi}$ by any of the other isomorphisms between vector fields and differential forms, but we omit these results for brevity with the hope that they are obvious.

\section{A geometric and Hamiltonian formulation of Maxwell's equations}
We now investigate the Hamiltonian structure of the macroscopic Maxwell equations in the context of split exterior calculus.

\subsection{Hamiltonian structure of the macroscopic Maxwell equations}
First, we review the Hamiltonian structure as presented in \cite{morrison_gauge_free_lifting}. Consider a general matter model prescribed by
\begin{equation}
	K[ \bm{E}, \bm{B} ] = \int_\Omega \mathcal{K}(\bm{x}, \bm{E}, \bm{B}, \nabla \bm{E}, \nabla \bm{B}, ... ) \vol^3
\end{equation}
so that the polarization and magnetization are defined by
\begin{equation}
	\bm{P}(\bm{x},t) = - \frac{\delta K}{\delta \bm{E}} \quad \text{and} \quad \bm{M}(\bm{x},t) = - \frac{\delta K}{\delta \bm{B}}.
\end{equation}
The Hamiltonian of the macroscopic Maxwell equations system is given by
\begin{equation}
	H[\bm{E},\bm{B}] = K - \int_\Omega \bm{E} \cdot \frac{\delta K}{\delta \bm{E}} \vol^3 + \frac{1}{8 \pi} \int_M ( \bm{E} \cdot \bm{E} + \bm{B} \cdot \bm{B} ) \vol^3.
\end{equation}
The Poisson bracket is defined to be
\begin{equation}
	\{ F, G \} = 4 \pi c \int_\Omega \left[ \frac{\delta F}{\delta \bm{D}} \cdot \nabla \times \frac{\delta G}{ \delta \bm{B}} 
				- \frac{\delta G}{\delta \bm{D}} \cdot \nabla \times \frac{\delta F}{\delta \bm{B}} \right] \vol^3.
\end{equation}
This bracket has Casimir invariants $\mathcal{C}_{\bm{D}} = \mathcal{F}[\nabla \cdot \bm{D}]$ and $\mathcal{C}_{\bm{B}} = \mathcal{F}[\nabla \cdot \bm{B}]$ where $\mathcal{F}$ is an arbitrary smooth functional. Hence, Gauss's law and the absence of magnetic monopoles hold as long as they are enforced in the initial conditions. 

In order to obtain the equations of motion, we need to take derivatives of the Hamiltonian with respect to $(\bm{D}, \bm{B})$. As the details of this procedure are omitted in \cite{morrison_gauge_free_lifting}, it is useful to show the full calculation here since we will perform the calculation again in the language of exterior calculus in the following section.
\begin{lemma}
If we think of $\bm{E}$ as an implicit function of $(\bm{D}, \bm{B})$, then
\begin{equation}
	\frac{\delta \bm{E}}{\delta \bm{D}} = \left( I - 4 \pi \frac{\delta^2 K}{\delta \bm{E} \delta \bm{E}} \right)^{-1}
	\quad \text{and} \quad
	\frac{\delta \bm{E}}{\delta \bm{B}} = 4 \pi \left( I - 4 \pi \frac{\delta^2 K}{\delta \bm{E} \delta \bm{E}}  \right)^{-1} \frac{\delta^2 K}{\delta \bm{B} \delta \bm{E}}.
\end{equation}
\end{lemma}
\noindent \textit{Proof:} Let $\Phi[ \bm{E}, \bm{B}] = (\bm{D}, \bm{B})$. That is,
\begin{equation*}
	\Phi[ \bm{E}, \bm{B}] = \left( \bm{E} - 4 \pi \frac{\delta K}{\delta \bm{E}}, \bm{B} \right).
\end{equation*}
We assume that $K$ is such that $\Phi$ is a diffeomorphism. Hence, we also have $\Phi^{-1}[ \bm{D}, \bm{B}] = (\bm{E}, \bm{B})$. We have that
\begin{equation*}
	\begin{aligned}
    		D \Phi [\bm{E},\bm{B}](\delta \bm{E}, \delta \bm{B}) 
    		&= 
		\begin{pmatrix}
			D_1 \Phi_1[\bm{E},\bm{B}] & D_2 \Phi_1[\bm{E},\bm{B}] \\
			0 & 1
		\end{pmatrix}
		\begin{pmatrix}
			\delta \bm{E} \\
			\delta \bm{B}
		\end{pmatrix} \\
		&=
		\begin{pmatrix}
			I - 4 \pi \frac{\delta^2 K}{\delta \bm{E} \delta \bm{E}} & - 4 \pi \frac{\delta^2 K}{\delta \bm{B} \delta \bm{E}} \\
			0 & 1
		\end{pmatrix}
		\begin{pmatrix}
			\delta \bm{E} \\
			\delta \bm{B}
		\end{pmatrix}.
	\end{aligned}
\end{equation*}
Hence, it follows that
\begin{equation*}
  	D \Phi^{-1}[\bm{D},\bm{B}](\delta \bm{D}, \delta \bm{B}) 
	= 
	\begin{pmatrix}
		D_1 \Phi_1[\bm{E},\bm{B}]^{-1} & - D_1 \Phi_1[\bm{E},\bm{B}]^{-1} D_2 \Phi_1[\bm{E},\bm{B}] \\
		0 & 1
	\end{pmatrix}
	\begin{pmatrix}
		\delta \bm{D} \\
		\delta \bm{B}
	\end{pmatrix}
\end{equation*}
If we compute the entries of this matrix, one finds
\begin{equation*}
 	\frac{\delta \bm{E}}{\delta \bm{D}} = \left( I - 4 \pi \frac{\delta^2 K}{\delta \bm{E} \delta \bm{E}}  \right)^{-1}
	\quad \text{and} \quad
	\frac{\delta \bm{E}}{\delta \bm{B}} = 4 \pi \left( I - 4 \pi \frac{\delta^2 K}{\delta \bm{E} \delta \bm{E}}  \right)^{-1} \frac{\delta^2 K}{\delta \bm{B} \delta \bm{E}}.
\end{equation*}
\qed

\begin{prop}
Let $\overline{H}[\bm{D}, \bm{B}] = H[\bm{E}, \bm{B}]$. Then
\begin{equation} \label{vec_ham_derivs}
	\frac{\delta \overline{H}}{\delta \bm{D}} = \frac{\bm{E}}{4 \pi} \quad \text{and} \quad \frac{\delta \overline{H}}{\delta \bm{B}} = \frac{\bm{H}}{4 \pi}.
\end{equation}
\end{prop}
\noindent \textit{Proof:} Taking derivatives of $H$ with respect to $(\bm{E}, \bm{B})$, we find
\begin{equation*}
	\frac{\delta H}{\delta \bm{E} } = \left( I - 4 \pi \frac{\delta K}{\delta \bm{E} \delta \bm{E}} \right) \frac{\bm{E}}{4 \pi} \quad \text{and} \quad
	\frac{\delta H}{\delta \bm{B} } = \frac{\delta K}{\delta \bm{B}} - \left( \frac{\delta K}{\delta \bm{B} \delta \bm{E}} \right)^* \bm{E} + \frac{\bm{B}}{4 \pi}.
\end{equation*}
The chain rule implies
\begin{equation*}
	\frac{\delta \overline{H}}{\delta \bm{D}} = \left( \frac{\delta \bm{E}}{\delta \bm{D}} \right)^* \frac{\delta H}{\delta \bm{E}} = \left( \frac{\delta \bm{E}}{\delta \bm{D}} \right)^* \left( I - 4 \pi \frac{\delta K}{\delta \bm{E} \delta \bm{E}} \right) \frac{\bm{E}}{4 \pi} = \frac{\bm{E}}{4 \pi}.
\end{equation*}
Likewise, 
\begin{align*}
	\frac{\delta \overline{H}}{\delta \bm{B}} 
	&= \frac{\delta H}{\delta \bm{B}} + \left( \frac{\delta \bm{E}}{\delta \bm{B}} \right)^* \frac{\delta H}{\delta \bm{E}} \\
	&= \frac{\delta K}{\delta \bm{B}} - \left( \frac{\delta^2 K}{\delta \bm{B} \delta \bm{E}} \right)^* \bm{E} + \frac{\bm{B}}{4 \pi} 
	+ \left( \frac{\delta \bm{E}}{\delta \bm{B}} \right)^* \left( I - 4 \pi \frac{\delta K}{\delta \bm{E} \delta \bm{E}} \right) \frac{\bm{E}}{4 \pi} \\
	&= \frac{\bm{B}}{4 \pi} + \frac{\delta K}{\delta \bm{B}} = \frac{\bm{H}}{4 \pi}.
\end{align*}
\qed

Hence, we find the evolution of an arbitrary observable. For $F = F[\bm{D}, \bm{B}]$,
\begin{equation}
	\dot{F} = \{F, \overline{H} \} = c \int_M \left[ \frac{\delta F}{\delta \bm{D}} \cdot \nabla \times \bm{H} - \bm{E} \cdot \nabla \times \frac{\delta F}{\delta \bm{B}} \right] \vol^3.
\end{equation}
\noindent Under the assumption of homogeneous boundary conditions,
\begin{equation*}
	\int_M U \cdot \nabla \times V \vol^3 = - \int_M \nabla \times U \cdot V \vol^3,
\end{equation*}
we immediately obtain the equations of motion:
\begin{equation}
	\frac{\partial \bm{D}}{\partial t} = \{ \bm{D}, H \} = c \nabla \times \bm{H} 
	\quad \text{and} \quad
	\frac{\partial \bm{B}}{\partial t} = \{ \bm{B}, H \} = - c \nabla \times \bm{E}.
\end{equation}
In this paper, Maxwell's equations have not been coupled to a dynamical model governing the source terms. Hence, because we are only considering self-consistent dynamics generated by a non-canonical Poisson bracket, i.e. we set $\rho_f = 0$ and $\bm{J}_f = 0$. One may augment the theory by adding, for example, a fluid or kinetic model for the source terms. 

\subsection{The Hamiltonian structure of the macroscopic Maxwell equations in split exterior calculus}
Recalling the manner of identifying the vector calculus variables with their exterior calculus counterparts given in section \ref{sec:macro_maxwell_1}, we may directly translate the Hamiltonian and Poisson bracket. The Hamiltonian may be written
\begin{equation} \label{geometric_hamiltonian}
	H[\bm{e}^1, \bm{b}^2] = K - \left( \bm{e}^1, \frac{\delta K}{\delta \bm{e}^1} \right) + \frac{1}{8 \pi} \left[ \left( \bm{e}^1, \bm{e}^1 \right) + \left( \bm{b}^2, \bm{b}^2 \right) \right],
\end{equation}
and the Poisson bracket may be written
\begin{equation} \label{pb_diff_forms}
	\{ F, G \} = 4 \pi c \left[ \left\langle \frac{\tilde{\delta} F}{\delta \tilde{\bm{d}}^2}, \mathsf{d} \frac{\tilde{\delta} G}{\delta \bm{b}^2} \right\rangle 
			- \left\langle \frac{\tilde{\delta} G}{\delta \tilde{\bm{d}}^2}, \mathsf{d} \frac{\tilde{\delta} F}{ \delta \bm{b}^2 } \right\rangle \right].
\end{equation}
As written, we clearly see that the Hamiltonian is metric dependent, being defined in terms of the $L^2$ inner product, while the Poisson bracket is explicitly metric free. 

To obtain the equations of motion, we need to take derivatives of the Hamiltonian. The following results are entirely analogous to those we found in the context of vector calculus, but are included to emphasize the equivalence of the two perspectives.
\begin{prop}
\begin{equation}
	\frac{\tilde{\delta} H}{\delta \bm{e}^1} = \left( \tilde{\star}_1 - 4 \pi \frac{\tilde{\delta} K}{\delta \bm{e}^1 \delta \bm{e}^1} \right) \frac{\tilde{\star}_1 \bm{e}^1}{4 \pi} \quad \text{and} \quad
	\frac{\tilde{\delta} H}{\delta \bm{b}^2} = \frac{\tilde{\delta} K}{\delta \bm{b}^2} - \left( \frac{\tilde{\delta}^2 K}{\delta \bm{b}^2 \delta \bm{e}^1} \right)^* \bm{e}^1 + \frac{ \tilde{\star}_2 \bm{b}^2}{4 \pi}.
\end{equation}
\end{prop}
\noindent \textit{Proof:} Taking the variation of $H$, we find
\begin{equation*}
	D_1 H[\bm{e}^1, \bm{b}^2] \delta \bm{e}^1 
	= \left\langle \delta \bm{e}^1, \frac{\tilde{\delta} K}{\delta \bm{e}^1} \right\rangle
		- \left\langle \delta \bm{e}^1, \frac{\tilde{\delta} K}{\delta \bm{e}^1} \right\rangle
		- \left\langle \bm{e}^1, \frac{\tilde{\delta}^2 K}{\delta \bm{e}^1 \delta \bm{e}^1} \delta \bm{e}^1 \right\rangle
		+ \frac{1}{4 \pi} \langle \delta \bm{e}^1, \tilde{\star}_1 \bm{e}^1 \rangle.
\end{equation*}
Hence, we find
\begin{equation*}
	\frac{\tilde{\delta} H}{\delta \bm{e}^1} = \left( \tilde{\star}_1 - 4 \pi \frac{\tilde{\delta} K}{\delta \bm{e}^1 \delta \bm{e}^1} \right) \frac{\tilde{\star}_1 \bm{e}^1}{4 \pi}.
\end{equation*}
Similarly,
\begin{equation*}
	D_2 H[\bm{e}^1, \bm{b}^2] \delta \bm{b}^2 
	= \left\langle \delta \bm{b}^2, \frac{\tilde{\delta} H}{\delta \bm{b}^2} \right\rangle 
	= \left\langle \delta \bm{b}^2, \frac{\tilde{\delta} K}{\delta \bm{b}^2} \right\rangle
	- \left\langle \delta \bm{e}^1, \frac{\tilde{\delta}^2 K}{\delta \bm{b}^2 \delta \bm{e}^1} \delta \bm{b}^2 \right\rangle 
	+ \frac{1}{4 \pi} \langle \delta \bm{b}^2, \tilde{\star}_2 \bm{b}^2 \rangle.
\end{equation*}
Hence,
\begin{equation*}
	\frac{\tilde{\delta} H}{\delta \bm{b}^2} = \frac{\tilde{\delta} K}{\delta \bm{b}^2} - \left( \frac{\tilde{\delta}^2 K}{\delta \bm{b}^2 \delta \bm{e}^1} \right)^* \bm{e}^1 + \frac{ \tilde{\star}_2 \bm{b}^2}{4 \pi}.
\end{equation*}
\qed

\noindent Equation \ref{eq:deriv_of_map}, allows us to conclude
\begin{equation*}
  \frac{\delta \bm{D}}{\delta \bm{E}}V = \left( I - 4 \pi \frac{\delta^2 K}{\delta \bm{E} \delta \bm{E}}\right) V \iff \frac{\tilde{\delta} \tilde{\bm{d}}^2}{\delta \bm{e}^1} v^1 = \left( \tilde{\star}_1 - 4 \pi \frac{\tilde{\delta}^2 K}{\delta \bm{e}^1 \delta \bm{e}^1} \right) v^1. 
\end{equation*}
Likewise, we similar reasoning finds
\begin{equation*}
  \frac{\delta \bm{E}}{\delta \bm{B}}V = 4 \pi \left( I - 4 \pi \frac{\delta^2 K}{\delta \bm{E} \delta \bm{E}}  \right)^{-1} \frac{\delta^2 K}{\delta \bm{B} \delta \bm{E}}V 
  \iff
  \frac{\tilde{\delta} \bm{e}^1}{\delta \bm{b}^2} \tilde{v}^2 = 4 \pi \left( \tilde{\star}_1 - 4 \pi \frac{\tilde{\delta}^2 K}{\delta \bm{e}^1 \delta \bm{e}^1} \right)^{-1} \frac{ \tilde{\delta}^2 K}{\delta \bm{b}^2 \delta \bm{e}^1} \tilde{v}^2. 
\end{equation*}

\begin{prop}
Define $\overline{H}[\tilde{\bm{d}}^2, \bm{b}^2] = H[\bm{e}^1, \bm{b}^2]$. Then one finds
\begin{equation}
	\frac{\tilde{\delta} \overline{H}}{\delta \tilde{\bm{d}}^2} = \frac{\bm{e}^1}{4 \pi} \quad \text{and} \quad \frac{\tilde{\delta} \overline{H}}{\delta \bm{b}^2} = \frac{\tilde{\star}_2 \bm{b}^2}{4 \pi} + \frac{\tilde{\delta} K}{\delta \bm{b}^2} = \frac{\tilde{\bm{h}}^1}{4 \pi}.
\end{equation}
\end{prop}
\noindent \textit{Proof:} The chain rule implies
\begin{equation*}
	\frac{\tilde{\delta} \overline{H}}{\delta \tilde{\bm{d}}^2} = \frac{\tilde{\delta} \bm{e}^1}{\delta \tilde{\bm{d}}^2} \frac{\tilde{\delta} H }{\delta \bm{e}^1} = \frac{\tilde{\delta} \bm{e}^1}{\delta \tilde{\bm{d}}^2} \left( \tilde{\star}_1 - 4 \pi \frac{\tilde{\delta} K}{\delta \bm{e}^1 \delta \bm{e}^1} \right) \frac{\bm{e}^1}{4 \pi}
\end{equation*}
and
\begin{equation*}
	\frac{\tilde{\delta} \overline{H}}{\delta \bm{b}^2} 
	= \frac{\tilde{\delta} H}{\delta \bm{b}^2} + \frac{\tilde{\delta} \bm{e}^2}{\delta \bm{b}^2} \frac{\tilde{\delta} H }{\delta \bm{e}^1} 
	= \frac{\tilde{\delta} K}{\delta \bm{b}^2} - \left( \frac{\tilde{\delta}^2 K}{\delta \bm{b}^2 \delta \bm{e}^1} \right)^* \bm{e}^1 + \frac{ \tilde{\star}_2 \bm{b}^2}{4 \pi} 
	+ \frac{\tilde{\delta} \bm{e}^2}{\delta \bm{b}^2} \left( \tilde{\star}_1 - 4 \pi \frac{\tilde{\delta} K}{\delta \bm{e}^1 \delta \bm{e}^1} \right) \frac{\bm{e}^1}{4 \pi}.
\end{equation*}
Hence, we find
\begin{equation*}
	\frac{\tilde{\delta} \overline{H}}{\delta \tilde{\bm{d}}^2} = \frac{\bm{e}^1}{4 \pi}
	\quad \text{and} \quad
	\frac{\tilde{\delta} \overline{H}}{\delta \bm{b}^2} = \frac{\tilde{\star}_2 \bm{b}^2}{4 \pi} + \frac{\tilde{\delta} K}{\delta \bm{b}^2} = \frac{\tilde{\bm{h}}^1}{4 \pi}.
\end{equation*} 
\qed \\

\noindent The above calculation allows us to write
\begin{equation}
	D \overline{H}[ \tilde{\bm{d}}^2, \bm{b}^2](\delta \tilde{\bm{d}}^2, \delta \bm{b}^2) 
	= \left\langle \frac{\bm{e}^1}{4 \pi}, \delta \tilde{\bm{d}}^2 \right\rangle 
	+ \left\langle \frac{\tilde{\bm{h}}^1}{4 \pi}, \delta \bm{b}^2 \right\rangle.
\end{equation}
Therefore, $(\bm{e}^1, \tilde{\bm{h}}^1)$ may be thought of as the response to variation of the electromagnetic field energy, $\overline{H}$. 

Let $F = F[\tilde{\bm{d}}^2, \bm{b}^2]$. Then the evolution of this arbitrary functional is given by
\begin{equation}  \label{weak_eom}
	\dot{F} 
	= \{ F, \overline{H} \} 
	= c \left[ \left\langle \frac{\tilde{\delta} F}{\delta \tilde{\bm{d}}^2}, \mathsf{d}\tilde{\bm{h}^1} \right\rangle
		- \left\langle \bm{e}^1, \mathsf{d} \frac{\tilde{\delta} F}{\delta \bm{b}^2} \right\rangle \right].
\end{equation}
If $F = F[\tilde{\bm{d}}^2]$, we obtain Amp{\`e}re's law
\begin{equation}
	\dot{F} = \{ F[\tilde{\bm{d}}^2], \overline{H} \} = c \left\langle \frac{\tilde{\delta} F}{\delta \tilde{\bm{d}}^2}, \mathsf{d}\tilde{\bm{h}^1} \right\rangle \implies \partial_t \tilde{\bm{d}}^2 = c \mathsf{d} \tilde{\bm{h}}^1, 
\end{equation}
and if $F = F[\bm{b}^2]$, then we obtain Faraday's law
\begin{equation}
	\dot{F} = \{ F[\bm{b}^2], \overline{H} \} = -c \left\langle \bm{e}^1, \mathsf{d} \frac{\tilde{\delta} F}{\delta \bm{b}^2} \right\rangle \implies \partial_t \bm{b}^2 = - c \mathsf{d} \bm{e}^1. 
\end{equation}
The constraint equations are given by the Casimir invariants of the Poisson bracket which we infer from equation (\ref{chain_rule_ext_deriv}): $\mathcal{C}_{\bm{b}^2} = F[\mathsf{d} \bm{b}^2]$ and $\mathcal{C}_{\tilde{\bm{d}}^2} = F[\mathsf{d} \tilde{\bm{d}}^2]$. These Casimir invariants imply the non-evolving constraints in system (\ref{eqn:split_maxwell}): that is, Gauss's law and the nonexistence of magnetic monopoles.

\subsection{On the correspondence between the vector calculus and exterior calculus formulation}
We will now examine the structure of the Poisson bracket for Maxwell's equations in local coordinates. To begin, recall that
\begin{equation*}
	\frac{\tilde{\delta} F}{\delta \tilde{\bm{d}}^2} = \left( \frac{\delta F}{\delta \bm{D}} \right)^\flat 
	\quad \text{and} \quad 
	\frac{\tilde{\delta} F}{\delta \bm{b}^2} = \left( \frac{\delta F}{\delta \bm{B}} \right)^\flat.
\end{equation*}
Hence,
\begin{equation}
	\frac{\delta F}{\delta \bm{D}}^i \frac{\partial}{\partial x^i} =  g^{ij} \frac{\delta F}{\delta D^j} \frac{\partial}{\partial x^i}
	\iff
	\frac{\tilde{\delta} F}{\delta \tilde{\bm{d}}^2} = \frac{\delta F}{\delta D^i} \mathsf{d} x^i.
\end{equation}
Hence, unsurprisingly, the coordinates of $\tilde{\delta} F / \delta \tilde{\bm{d}}^2$ are just the components of the covector representation of $\delta F/ \delta \bm{D}$. The coordinate expression for $\tilde{\delta} F / \delta \bm{b}^2$ is analogous. 

\begin{prop}
Using the above coordinate expressions, we find that
\begin{equation}
	\{F, G\} = 4 \pi c \int_M \left( \frac{\delta F}{\delta D^i}  \frac{\partial}{\partial x^j} \left( \frac{\delta G}{\delta B^k} \right) - \frac{\delta \mathcal{G}}{\delta D^i}  \frac{\partial}{\partial x^j} \left( \frac{\delta \mathcal{F}}{\delta B^k} \right) \right) \mathsf{d} x^i \wedge \mathsf{d} x^j \wedge \mathsf{d} x^k.
\end{equation}
\end{prop}
\noindent \textit{Proof:}
\begin{align*}
	\{F, G\} &= 4 \pi c \left[ \left\langle \frac{\delta F}{\delta D^i} \mathsf{d} x^i, \mathsf{d} \left( \frac{\delta G}{\delta B^k} \mathsf{d} x^k \right) \right\rangle_{\Lambda^1, \tilde{\Lambda}^2} - \left\langle \frac{\delta G}{\delta D^i} \mathsf{d} x^i, \mathsf{d} \left( \frac{\delta F}{\delta B^k} \mathsf{d} x^k \right) \right\rangle_{\Lambda^1, \tilde{\Lambda}^2} \right] \\
	&= 4 \pi c \left[ \left\langle \frac{\delta F}{\delta D^i} \mathsf{d} x^i, \frac{\partial}{\partial x^j} \frac{\delta G}{\delta B^k} \mathsf{d} x^j \wedge \mathsf{d} x^k \right\rangle_{\Lambda^1, \tilde{\Lambda}^2} - \right. \\
	&\hspace{5cm} \left. - \left\langle \frac{\delta G}{\delta D^i} \mathsf{d} x^i, \frac{\partial}{\partial x^j} \frac{\delta F}{\delta B^k} \mathsf{d} x^j \wedge \mathsf{d} x^k \right\rangle_{\Lambda^1, \tilde{\Lambda}^2} \right] \\
	& = 4 \pi c \int_M \left( \frac{\delta F}{\delta D^i}  \frac{\partial}{\partial x^j} \left( \frac{\delta G}{\delta B^k} \right) - \frac{\delta \mathcal{G}}{\delta D^i}  \frac{\partial}{\partial x^j} \left( \frac{\delta \mathcal{F}}{\delta B^k} \right) \right) \mathsf{d} x^i \wedge \mathsf{d} x^j \wedge \mathsf{d} x^k.
\end{align*}
\qed

\noindent This coordinate representation makes it clear that the Poisson bracket has no metric dependence. As the Poisson bracket is a purely topological entity, that the bracket is expressed without reference to the metric is favorable. 

We next explicitly demonstrate the equivalence of the vector calculus and exterior calculus formulations of the Poisson bracket.
\begin{prop}
The Poisson bracket for the exterior calculus formulation is equivalent to that of the vector calculus formulation:
\begin{align*}
	\{F, G\} 
		&= 4 \pi c \left[ \left\langle \frac{\tilde{\delta} F}{\delta \tilde{\bm{d}}^2}, \mathsf{d} \frac{\tilde{\delta} G}{\delta \bm{b}^2} \right\rangle
		 	- \left\langle \frac{\tilde{\delta} G}{\delta \tilde{\bm{d}}^2}, \mathsf{d} \frac{\tilde{\delta} F}{ \delta \bm{b}^2 } \right\rangle \right] \\
		&= 4 \pi c \int_M \left[ \frac{\delta F}{\delta \bm{D}} \cdot \nabla \times \frac{\delta G}{\delta \bm{B}} 
			- \frac{\delta G}{\delta \bm{D}} \cdot \nabla \times \frac{\delta F}{\delta \bm{B}} \right] \vol^3. 
\end{align*}
\end{prop}
\noindent \textit{Proof:} We have that
\begin{equation*}
	\frac{\tilde{\delta} F}{\delta \tilde{\bm{d}}^2} = \left( \frac{\delta F}{\delta \bm{D}} \right)^\flat
	\implies
	\mathsf{d} \frac{\tilde{\delta} F}{\delta \tilde{\bm{d}}^2}  = \textbf{i}_{\nabla \times \frac{\delta F}{\delta \bm{D}}} \vol^3. 
\end{equation*}
Similar results hold for the functional derivatives with respect $\bm{b}^2$. Hence, we find
\begin{equation*}
	\frac{\tilde{\delta} F}{\delta \tilde{\bm{d}}^2} \wedge \mathsf{d} \frac{\tilde{\delta} G}{\delta \bm{b}^2} = \frac{\delta F}{\delta \bm{D}} \cdot \nabla \times \frac{\delta G}{\delta \bm{B}} \vol^3.
\end{equation*}
From this follows the result. \qed 

We find that the Poisson bracket satisfies the Jacobi identity. If we use the notation
\begin{align*}
	DF[\tilde{\bm{d}}^2, \bm{b}^2](\tilde{u}^2, v^2) &= D_1F[\tilde{\bm{d}}^2, \bm{b}^2] \tilde{u}^2 + D_2 F[\tilde{\bm{d}}^2, \bm{b}^2] v^2 \\
	&= \left\langle \frac{\tilde{\delta} F}{\delta \tilde{\bm{d}}^2}, \tilde{u}^2 \right\rangle + \left\langle \frac{\tilde{\delta} F}{\delta \bm{b}^2}, v^2 \right\rangle 
	=: \left\langle \frac{\tilde{\delta} F}{\delta (\tilde{\bm{d}}^2, \bm{b}^2) }, (\tilde{u}^2, v^2) \right\rangle
\end{align*}
then the Poisson bracket may be written
\begin{equation*}
	\{F, G\} =  \left\langle \frac{\tilde{\delta} F}{\delta (\tilde{\bm{d}}^2, \bm{b}^2) }, 
	\begin{pmatrix}
		0 & \mathsf{d} \\
		- \mathsf{d} & 0
	\end{pmatrix}
	\frac{\tilde{\delta} G}{\delta (\tilde{\bm{d}}^2, \bm{b}^2)} \right\rangle.
\end{equation*}
The application of the Poisson tensor written in this notationally suggestive matrix format should be contextually clear. This way of writing the Poisson bracket makes it entirely clear that the Poisson tensor is independent of the fields. Hence, it follows that antisymmetry is sufficient for the Jacobi identity to hold \cite{morrison_poisson_brackets}. 

\subsection{Expressing the system with a single de Rham complex}

As a final note, we may formulate the entire theory in terms of only the primal de Rham complex of straight differential forms. This way of expressing the equations obfuscates the metric independence of the Poisson bracket, and requires use of the codifferential operator \cite{arnold_FEEC}. This perspective is complementary to the double de Rham complex formulation, and aids in providing a complete understanding of the mathematical structure of the model. Further, this approach expresses the equations in a formate amenable to discretization by traditional finite element methods whereas the split exterior calculus modeling paradigm in this article is suited to discretization strategy which explicitly constructs distinct finite element spaces for the primal and dual de Rham complexes \cite{dualMimeticMaxwell}.

The Hamiltonian remains the same as before, but the Poisson bracket is written
\begin{equation} \label{pb_straight_diff_forms}
	\{ F, G \} = 4 \pi c \left[ \left( \frac{\delta F}{\delta \bm{d}^1}, \mathsf{d}^* \frac{\delta G}{\delta \bm{b}^2} \right) - \left( \frac{\delta G}{\delta \bm{d}^1}, \mathsf{d}^* \frac{\delta F}{ \delta \bm{b}^2 } \right) \right]
\end{equation}
and the constitutive laws are given by
\begin{equation}
	\bm{d}^1 = \bm{e}^1 - 4 \pi \frac{\delta K}{\delta \bm{e}^1} \quad \text{and} \quad \bm{h}^2 = \bm{b}^2 + 4 \pi \frac{\delta K}{\delta \bm{b}^2}.
\end{equation}
These are obtained from the split exterior calculus formulation simply by noting the relationship between the twisted and straight functional derivatives, and the relationship between the $L^2$ inner product and the Poincar{\'e} duality pairing. Because
\begin{equation}
	\frac{\delta \overline{H}}{\delta \bm{d}^1} = \frac{\bm{e}^1}{4 \pi} \quad \text{and} \quad \frac{\delta \overline{H}}{\delta \bm{b}^2} = \frac{ \bm{b}^2}{4 \pi} + \frac{\delta K}{\delta \bm{b}^2} = \frac{\bm{h}^2}{4 \pi}
\end{equation}
it follows that for any functional $F = F[\bm{d}^1, \bm{b}^2]$,
\begin{equation} \label{weak_straight_eom}
	\begin{split}
		\dot{F} 
			= \{ F, \overline{H} \} 
			= c \left[ \left( \frac{ \delta F}{\delta \bm{d}^1}, \mathsf{d}^* \bm{h}^2 \right) 
				- \left( \bm{e}^1, \mathsf{d}^* \frac{\delta F}{\delta \bm{b}^2} \right) \right] 
	\end{split}
	\quad \implies \quad
	\begin{split}
		\partial_t \bm{d}^1 &= c \mathsf{d}^* \bm{h}^2 \\
		\partial_t \bm{b}^2 &= - c \mathsf{d} \bm{e}^1.
	\end{split}
\end{equation}
Hence, one can see that, with respect to this formulation of the macroscopic Maxwell equations, Ampere's law must be expressed weakly while Faraday's law is expressed strongly. If $F[\omega] = \hat{F}[\mathsf{d} \omega]$ ( resp. $F[\omega] = \hat{F}[\mathsf{d}^* \omega]$), then
\begin{equation*}
	\frac{\delta F}{\delta \omega} = \mathsf{d}^* \frac{\delta \hat{F}}{\delta \mathsf{d} \omega} 
	\quad \left( \text{resp.} \quad 
	\frac{\delta F}{\delta \omega} = \mathsf{d} \frac{\delta \hat{F}}{\delta \mathsf{d}^* \omega} \right).
\end{equation*}
Therefore, it follows that the Poisson bracket has Casimir invariants $F[ \mathsf{d}^* \bm{d}^1]$ and $F[ \mathsf{d} \bm{b}^2]$ for arbitrary smooth functional $F$. Hence, the no-magnetic-monopole condition is strongly imposed while charge conservation is weakly imposed. 

Finally, we could reduce to a single de Rham complex by instead eliminating all of the straight forms and expressing the theory entirely in terms of twisted forms. Doing this, we would find that Ampere's law and charge conservation are strongly imposed while Faraday's law and the exclusion of magnetic monopoles are weakly imposed. 

\subsection{Example matter models}
We conclude by briefly considering three simple constitutive models that might be accommodated by this framework. This is done to hint at the breadth of constitutive models that might be self consistently defined, and that a well defined Hamiltonian accompanies each model. 

First, consider an intensity dependent index of refraction:
\begin{equation}
	\bm{P} = \left( \chi^1 + \chi^3 | \bm{E} |^2 \right) \bm{E} \iff \bm{p}^1 = \left( \chi^1 + \chi^3 | \bm{e}^1 |^2 \right) \bm{e}^1
\end{equation}
where $\chi^1$ and $\chi^3$ are scalars for simplicity. Such a model accounts for the lowest order nonlinear effects found in noncentrosymmetric media, and has been used to account for laser self-focusing in plasmas (see e.g. \cite{BoydNLO} and \cite{ShenNLO}). The $K$ functional leading to this polarization is
\begin{equation}
	K = - \int_\Omega \left( \frac{\chi^1}{2} | \bm{E} |^2 +  \frac{\chi^3}{4} | \bm{E} |^4 \right) \vol \iff K = - \int_\Omega \left( \frac{\chi^1}{2} | \bm{e}^1 |^2 + \frac{\chi^3}{4} | \bm{e}^1 |^4 \right) \vol.
\end{equation}
We find that the Hamiltonian of such a system is given by
\begin{equation}
	\begin{aligned}
		H &= \frac{1}{8 \pi} \int_\Omega \left[ \left( 1 + 4 \pi \chi^1 \right) | \bm{E} |^2 + | \bm{B} |^2 + 6 \pi \chi^3 | \bm{E} |^4 \right] \vol \\
			&= \frac{1}{8 \pi} \int_\Omega \left[ \left( 1 + 4 \pi \chi^1 \right) | \bm{e}^1 |^2 + | \bm{b}^2 |^2 + 6 \pi \chi^3 | \bm{e}^1 |^4 \right] \vol.
	\end{aligned}
\end{equation}
We could proceed in an analogous manner for the magnetic field.

Next, we might consider a system where the polarization depends on the electric field nonlocally in space. For example,
\begin{equation}
	\bm{P} = \alpha \bm{E} + \beta \Delta \bm{E} = \alpha \bm{E} + \beta \left( \nabla ( \nabla \cdot \bm{E}) - \nabla \times \nabla \times \bm{E} \right)
\end{equation}
or in terms of (straight) differential forms,
\begin{equation}
	\bm{p}^1 = \alpha \bm{e}^1 + \beta \left( \mathsf{d} \mathsf{d}^* + \mathsf{d}^* \mathsf{d} \right) \bm{e}^1.
\end{equation}
In this case, we find
\begin{equation}
	\begin{aligned}
		K &= - \frac{1}{2} \int_\Omega \left[ \alpha | \bm{E} |^2 + \beta \left( | \nabla \cdot \bm{E} |^2 + | \nabla \times \bm{E} |^2 \right) \right] \vol \\
			&= - \frac{1}{2} \int_\Omega \left[ \alpha | \bm{e}^1 |^2 + \beta \left( | \mathsf{d}^* \bm{e}^1 |^2 + | \mathsf{d} \bm{e}^1 |^2 \right) \right] \vol
	\end{aligned}
\end{equation}
Hence, assuming homogeneous boundary conditions, one obtains the Hamiltonian
\begin{equation}
	\begin{aligned}
		H &= \frac{1}{8 \pi} \int_\Omega \left[ \left(1 + 4 \pi \alpha \right) | \bm{E} |^2 + | \bm{B} |^2
			+ 4 \pi \beta \left( | \nabla \cdot \bm{E} |^2 + | \nabla \times \bm{E} |^2 \right) \right] \vol \\
		&= \frac{1}{8 \pi} \int_\Omega \left[ \left(1 + 4 \pi \alpha \right) | \bm{e}^1 |^2 + | \bm{b}^2 |^2
			+ 4 \pi \beta \left( | \mathsf{d}^* \bm{e}^1 |^2 + | \mathsf{d} \bm{e}^1 |^2 \right) \right] \vol.
	\end{aligned}
\end{equation}
This yields a Maxwell wave equation of the form
\begin{equation}
	\partial_t^2 \left[ ( \alpha + \beta \Delta )^{-1} \bm{E} \right] + c^2 \nabla \times \nabla \times \bm{E} = 0.
\end{equation}
Restricting our attention temporarily to 1D plane wave solutions of Maxwell's equations, we obtain the dispersion relation
\begin{equation}
	\omega(k) = \pm \sqrt{ \frac{ c^2 k^2}{ \alpha + \beta k^2 }} \implies v_g(k) = \pm \frac{\alpha}{c^2} \left( \frac{c^2}{\alpha + \beta k^2} \right)^{3/2}, \ v_{ph}(k) = \pm \sqrt{ \frac{c^2}{\alpha + \beta k^2}}.
\end{equation}
Hence, this Hamiltonian models a dispersive medium which retards the propagation of high wavenumber modes. We could proceed in an analogous manner to define a magnetization with nonlocal dependence on the magnetic field. 

It is also possible to define models with polarization dependent on the magnetic field and magnetization dependent on the electric field:
\begin{equation}
	K = - \frac{\alpha}{2} \int_\Omega | \bm{B} |^2 | \bm{E} |^2 \vol \implies \bm{P} = \alpha | \bm{B} |^2 \bm{E} \quad \text{and} \quad \bm{M} = \alpha | \bm{E} |^2 \bm{B}
\end{equation}
or in terms of differential forms
\begin{equation}
	K = - \frac{\alpha}{2} \int_\Omega | \bm{b}^2 |^2 | \bm{e}^1 |^2 \vol \implies \bm{P} = \alpha | \bm{b}^2 |^2 \bm{e}^1 \quad \text{and} \quad \bm{M} = \alpha | \bm{e}^1 |^2 \bm{b}^2.
\end{equation}
The Hamiltonian is given by
\begin{equation}
	\begin{aligned}
		H &= \frac{1}{8 \pi} \int_\Omega \left( | \bm{E} |^2 + 4 \pi \alpha | \bm{E} |^2 | \bm{B} |^2 + | \bm{B} |^2 \right) \vol \\
		&= \frac{1}{8 \pi} \int_\Omega \left( | \bm{e}^1 |^2 + 4 \pi \alpha | \bm{e}^1 |^2 | \bm{b}^2 |^2 + | \bm{b}^2 |^2 \right) \vol.
	\end{aligned}
\end{equation}
When we write $K$ in terms of differential forms, the consistency requirements of how products of differential forms might be formed impose restrictions on how we can mix the electric and magnetic fields in $K$. For example, $\bm{e}^1 \wedge \bm{b}^2$, while being a $3$-form, is not a density. Therefore, its integral is orientation dependent which is undesirable for an energy functional. Hence, this modeling framework helps to uncover subtle inconsistencies in complex constitutive relations.

\section{Conclusion}
This work brings together two distinct modeling formats for Maxwell's equations: its Hamiltonian formulation, and its formulation using the double de Rham complex. This work provides a theory for the macroscopic Maxwell equations which is flexible in specifying a broad class of polarizations and magnetizations; has Hamiltonian structure; is stated in the language of exterior calculus in a coordinate free, geometric language; is independent of the orientation of ambient space; and separates the portions of the theory which are metric independent (the Poisson bracket) and metric dependent (the Hamiltonian). 

The use of the double de Rham complex allows us to carefully account for orientation and metric dependence. Our models are orientation independent because the Hamiltonians and Poisson brackets presented in this paper are written as the integral of a density (a twisted top form). Regarding metric independence in the Poisson bracket, the benefit of using Poincar{\'e} duality as opposed to $L^2$ duality seen in comparing (\ref{pb_diff_forms}) and (\ref{pb_straight_diff_forms}). The former makes no use of the metric while the later obfuscates this metric independence due to the presence of the $L^2$ inner product and codifferential operator. As the Poisson bracket is a purely topological quantity, the metric free formulation best captures its essential character. However, it is worth noting that the two forms are entirely equivalent, and that the formulation in terms of a single de Rham complex, is sometimes advantageous. Moreover, it is the case that the functionals the bracket generally acts on, e.g. the Hamiltonian, often contain metric dependence. 

Since $K$ generally depends on the metric, it follows that metric dependence is generally present in the mapping from $(\bm{D}, \bm{B})$ to $(\bm{E}, \bm{H})$. Further, as seen in \cite{morrison_gauge_free_lifting}, the simplest expression for the Poisson bracket is given when $(\bm{D}, \bm{B})$ are taken to be the evolving quantities, and $(\bm{E}, \bm{H})$ are considered to be derived quantities which are only computed when necessary. Because the Lorenz force law depends on $(\bm{E},\bm{B})$, it follows that coupling to a particle model would at least require evaluation of $\bm{E}(\bm{D}, \bm{B})$. In order to model a greater variety of phenomena, future work considering a greater variety of functionals $K$ (in particular choices of $K$ which give rise to more complicated nonlinear constitutive relations) and coupling to Hamiltonian models for charged particles is needed. 

It is further worth noting that the model considered here, the three-dimensional macroscopic Maxwell equations, may also be expressed in terms of split exterior calculus in Hamiltonian form in one-dimensional and two-dimensional variants. Indeed, the model presented does not explicitly require that the $n=3$. It is straightforward to generalize this model to Hodge wave equations with general constitutive relations in Hamiltonian form in arbitrary dimensions. The modeling paradigm given here may be applied to nonlinear wave equations of a very general kind in arbitrary dimensions. 

Beyond the many theoretical benefits of a model possessing Hamiltonian structure (e.g. see \cite{MorrisonP.J1998Hdot}, \cite{nonlinear_stability_of_fluid_and_plasma_equilibria}), there is great value in leveraging Hamiltonian structure to design structure preserving discretizations. The geometric electromagnetic particle-in-cell method (GEMPIC) \cite{GEMPIC} is a Poisson integrator developed for the Maxwell-Vlasov equations. Such a numerical scheme has built in stability and does not introduce unphysical dissipation. The goal of developing GEMPIC-like discretizations for more general kinetic models models of the kind described in \cite{morrison_gauge_free_lifting} and, in particular, a Hamiltonian gyrokinetic model \cite{burby_et_al_gyrokinetics} is a major motivation behind the geometric treatment of the macroscopic Maxwell equations presented in this paper. Structure preserving spatial discretization often relies on the structure of the de Rham complex of differential forms (e.g. see \cite{hiptmair_maxwell_equations}, \cite{arnold_FEEC}, and \cite{kreeft2011mimetic}). This in turn requires expressing the equations of motion in terms of differential forms. Hence, this this work studies the geometric structure of the macroscopic Maxwell equations as a preliminary step towards incorporating the lifted kinetic models from \cite{morrison_gauge_free_lifting} into a computational framework like that found in \cite{GEMPIC}. The paper \cite{dualMimeticMaxwell} proposes a structure preserving discretization of the Hamiltonian formulation of the macroscopic Maxwell equations presented in this paper.

\section{Acknowledgements}
We gratefully acknowledge the support of U.S. Dept. of Energy Contract \# DE-FG05-80ET-53088, NSF Graduate Research Fellowship \# DGE-1610403, and the Humboldt foundation.


\bibliographystyle{plain} 
\bibliography{geo_maxwell} %

\end{document}